\documentclass[3p]{elsarticle}

\usepackage{hyperref}
\usepackage{graphicx}
\usepackage[cmex10]{amsmath}
\usepackage{amssymb}
\begin{document}

\begin{frontmatter}

\title{Parallel loop cluster quantum Monte Carlo simulation of quantum magnets based on global union-find graph algorithm}

\author[phys,issp]{Synge Todo}
\author[issp]{Haruhiko Matsuo}
\author[fujitsu]{Hideyuki Shitara}

\address[phys]{Department of Physics, The University of Tokyo, Tokyo 113-0033, Japan}
\address[issp]{Institute for Solid State Physics, The University of Tokyo, Kashiwa 277-8581, Japan}
\address[fujitsu]{Application Research \& Development Div., Next Generation Technical Computing Unit, Fujitsu Limited, Kawasaki 211-8588, Japan}

\begin{abstract}
A large-scale parallel loop cluster quantum Monte Carlo simulation is presented.
On 24,576 nodes of the K computer, one loop cluster Monte Carlo update of the world-line configuration of the $S=1/2$ antiferromagnetic Heisenberg chain with $2.6 \times 10^6$ spins at inverse temperature $3.1 \times 10^5$ is executed in about 8.62 seconds, in which global union-find cluster identification on a graph of about 1.1 trillion vertices and edges is performed.
By combining the nonlocal global updates and the large-scale parallelization, we have virtually achieved about $10^{13}$-fold speed-up from the conventional local update Monte Carlo simulation performed on a single core.
We have estimated successfully the antiferromagnetic correlation length and the magnitude of the first excitation gap of the $S=4$ antiferromagnetic Heisenberg chain for the first time as $\xi = 1.040(7) \times 10^4$ and $\Delta = 7.99(5) \times 10^{-4}$, respectively.
\end{abstract}


\end{frontmatter}

\section{Introduction}
\label{sec:Introduction}

The study of strongly-correlated quantum systems is the foremost area of research in contemporary statistical and condensed-matter physics~\cite{AvellaM2011}, where the computational approaches are of increasing importance during recent years.
From the viewpoint of the computational physics for the quantum lattice models, remaining challenges are: supersolid in frustrated spin/bosonic lattice models, where diagonal and offdiagonal long-range orders coexist~\cite{YamamotoTM2009}; cold atoms on optical lattice, in which one can compare experiments with computations directly~\cite{HuoZCTS2011}; deconfined criticality, a direct continuous quantum phase transition between long-range ordered phase with incompatible symmetries~\cite{SenthilVBSF2004,HaradaSOMLWTK2013,SuzukiHMTK2015}; long-range and strongly anisotropic interactions, in which one observes effective reduction of spatial dimensions, exotic boundary effects, etc~\cite{YasudaTHAKTT2005,YasudaST2015,HoritaST2017}.
In order to tackle such fundamental and essential problems in strongly-correlated quantum systems, numerical simulations on large lattices are generally required, as the correlation length sometimes reaches millions of lattice constants and in many cases it indeed diverges due to strong fluctuations.
Demands on unbiased and efficient simulation algorithms thus become stronger and stronger in recent years.  

The quantum Monte Carlo method is one of the most promising tools as in principle it can simulate rather large lattices in any dimensions in {\em statistically exact} ways~\cite{Suzuki1994,LandauB2014,GubernatisKW2016}.
However, it is widely known that the conventional quantum Monte Carlo method based on local updates of world lines suffers from several drawbacks; ergodicity problem, fine-mesh slowing down, etc.
Especially, in the vicinity of the criticality the autocorrelation time in the Markov chain diverges as $\xi^z$, where $\xi$ is the correlation length and $z \simeq 2$ is the dynamical exponent.
This is called the {\em critical slowing down}.
The loop algorithm invented in 1993~\cite{EvertzLM1993,WieseY1994} and its extensions solve (or at least reduce) most of the drawbacks in the conventional method~\cite{Evertz2003,KawashimaH2004,Todo2012}.

The loop algorithm, which is a kind of cluster algorithm, realizes updates of world-line configuration by flipping nonlocal objects, called {\em loops}. It has been shown that it is fully ergodic and drastically reduces the autocorrelation time, often by orders of magnitude, especially at low temperatures. Furthermore, by using the continuous-time version of the algorithm, one can completely eliminate the discretization error originating from the Suzuki-Trotter decomposition; simulations can be performed directly in the so-called Trotter limit.

In high-performance computing, the importance of the non-floating-point operations (e.g.,
graph algorithms) has been widely
realized these days, especially in the field of data-intensive
applications~\cite{Graph500}.  Here, we present the world's first
peta-scale cluster algorithm quantum Monte Carlo simulation on the K
computer based on global union-find algorithm on a graph of about 1.1
trillion vertices and edges, solving the fundamental problems in
statistical and condensed-matter physics.

The organization of the present paper is as follows.
After introducing the loop cluster algorithm, which is based on the nonlocal updates of world-line configuration by using the union-find graph algorithm, in Sec.~\ref{sec:LoopAlgorithm}, we discuss in detail how effectively the parallel graph algorithm is implemented using the OpenMP-MPI hybrid parallelization in Sec.~\ref{sec:Parallelization}.
By using the highly parallelized loop cluster quantum Monte Carlo algorithm, we demonstrate in Sec.~\ref{sec:PerformanceAnalysis} that one Monte Carlo update of the world-line configuration of the $S=1/2$ antiferromagnetic Heisenberg chain with $2.6 \times 10^6$ spins at inverse temperature $3.1 \times 10^5$ is executed in about 8.62 seconds on 24,576 nodes of the K computer.
By combining the nonlocal cluster updates and the large-scale parallelization on the K computer, we have virtually achieved about $10^{13}$-fold speed-up.
In Sec.~\ref{sec:SimulationResults}, we represent the result of the quantum Monte Carlo simulation of the $S=4$ antiferromagnetic Heisenberg chain.
We have estimated successfully the antiferromagnetic correlation length and the magnitude of the first excitation gap for the first time as $1.040(7) \times 10^4$ and $7.99(5) \times 10^{-4}$, respectively.
Finally, conclusions and some remarks are presented in Sec.~\ref{sec:Conclusion}.

\section{The Loop Algorithm}
\label{sec:LoopAlgorithm}

We consider the antiferromagnetic Heisenberg model on a chain lattice
of $L$ sites.  The Hamiltonian is given by
\begin{align}
  \label{eqn:Hamiltonian}
  {\cal H} = J \sum_{j=1}^L {\bf S}_j \cdot {\bf S}_{j+1},
\end{align}
where the coupling constant $J>0$, and ${\bf S}_j = (S_j^x,S_j^y,S_j^z)$ is the quantum spin operator at site $j$ with spin length $S$ satisfying the standard commutation relations, e.g., $[S_j^x,S_j^y] = i \hbar S_j^z$.
Hereafter, we set $J=1$ and assume periodic boundary conditions, ${\bf S}_{j+L} = {\bf S}_j$.
We consider the case with $S=1/2$ for a while.
An extension to the higher-spin cases will be discussed in Sec.~\ref{sec:SimulationResults}.

The expectation value of an observable $A$ is given by
\begin{align}
  \langle A \rangle = {\rm Tr} \, A \exp (-\beta {\cal H}) / Z,
\end{align}
where $Z$ is the partition function
\begin{align}
  Z = {\rm Tr} \, \exp (-\beta {\cal H}),
\end{align}
$\beta = 1/k_{\rm B}T$ the inverse temperature, $T$ the temperature, and $k_{\rm B}$ the Boltzmann constant.
The density matrix $\exp(-\beta{\cal H})$ is regarded as an imaginary time evolution operator.
Since the Hamiltonian ${\cal H}$ as well as the observable $A$ are $2^L \times 2^L$ matrices, the numerical diagonalization is feasible only for $L \lesssim 40$.

\begin{figure}[tb]
\centerline{\includegraphics[width=4in]{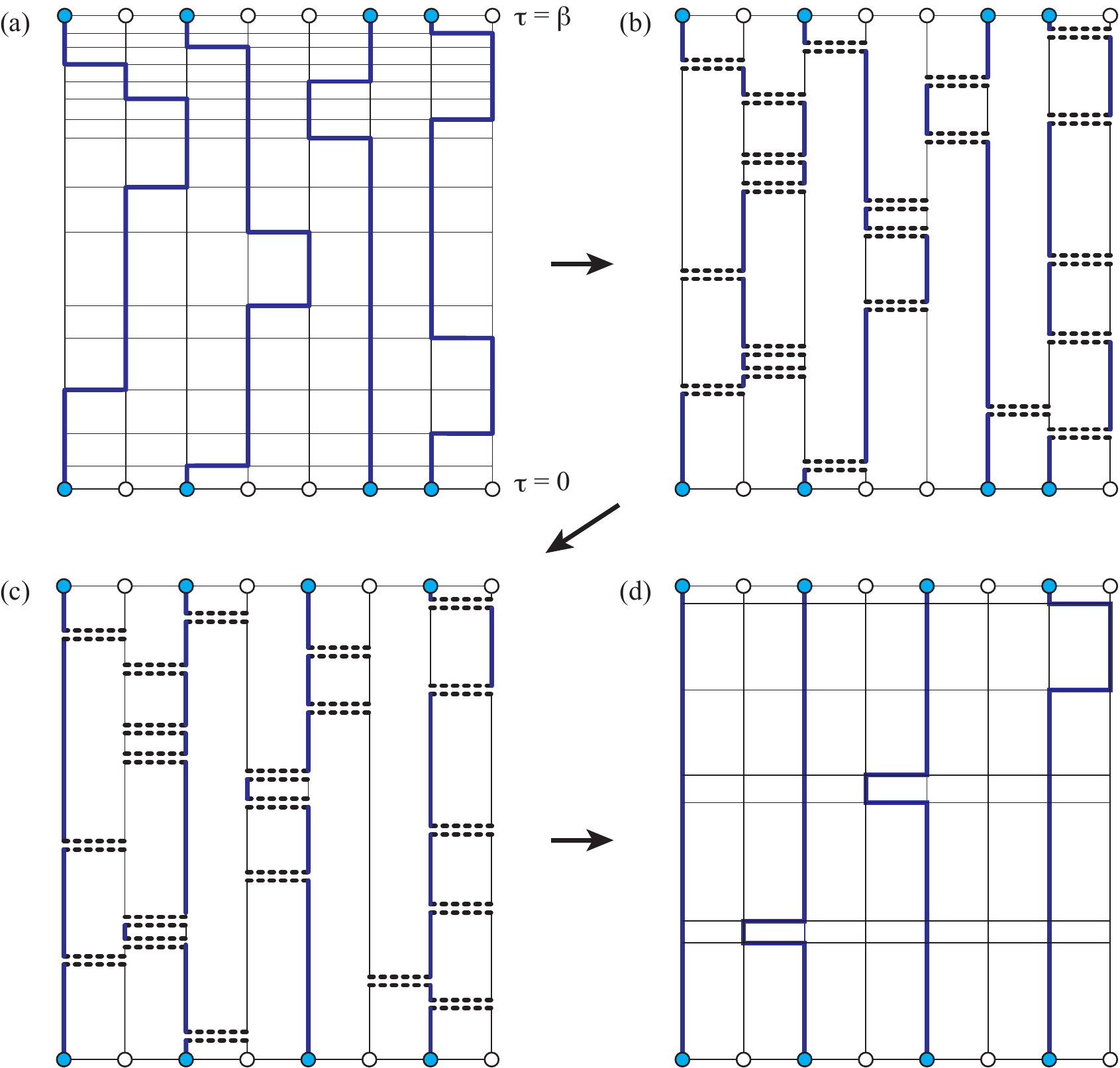}}
\caption{Example of one Monte Carlo step of the loop update for the antiferromagnetic Heisenberg chain with $L=8$.
  The horizontal and vertical axes correspond to the spatial and imaginary time directions, respectively.
  In (a) and (d), bold blue lines denote the world lines of up spins.}
\label{fig:loop}
\end{figure}

In the following, we work in a basis set $\{ | \phi \rangle \}$ in
which $S_j^z$ ($j=1 \cdots L$) are diagonalized, i.e., $| \phi \rangle
= | s_1, \cdots s_L \rangle$ with $s_j = \uparrow,\downarrow$.  We
first split the density matrix into $M$ time slices:
\begin{align}
  Z = {\rm Tr} \, [\exp (-\Delta_\tau {\cal H})]^M = {\rm Tr} \, {\cal T}^M +
  {\cal O}(\beta^2/M),
\end{align}
where $\Delta_\tau = \beta/M$ and we have introduced the quantum
transfer matrix ${\cal T}=1-\Delta_\tau {\cal H}$.  After inserting ($M-1$)
sums over a complete set of states, $\sum | \phi \rangle \langle \phi
| = 1$, between every ${\cal T}$, we obtain a path integral representation of
the partition function:
\begin{align}
  Z &\approx \! \! \sum_{\phi_M,\cdots,\phi_1} \! \! \langle \phi_1 |
  {\cal T} | \phi_M \rangle \langle \phi_M | {\cal T} | \phi_{M-1} \rangle \cdots
  \langle \phi_2 | {\cal T} | \phi_1 \rangle.
  \label{eq:part}
\end{align}
In the quantum Monte Carlo simulation, we sample the terms in the r.h.s.\ of Eq.~(\ref{eq:part}) according to their magnitude.
The matrix element $\langle \phi | {\cal T} | \phi' \rangle$ is nonzero if
and only if $|\phi\rangle = |\phi'\rangle$, or they are identical
except two neighboring spins that are swapped with one another, e.g.,
$|\phi\rangle = |\cdots\!\uparrow\downarrow\!\cdots\rangle$ and
$|\phi'\rangle =
|\cdots\!\downarrow\uparrow\!\cdots\rangle$.  These
constraints for configurations can be depicted in terms of {\em world
  lines} [Fig.~\ref{fig:loop}(a)].  Note that the diagonal elements in
${\cal T}$ are of ${\cal O} (1)$, while the nonzero offdiagonal elements are
of ${\cal O}(\Delta_\tau)$.  Therefore, the number of {\em jumps} of
world lines in the path integral representation remains finite even in
the Trotter limit $M \rightarrow \infty$, and it is enough to store
the spin configuration at $\tau = 0$ and the space-time positions of
jumps.

In the loop algorithm~\cite{EvertzLM1993}, given a world-line
configuration, we first divide it into a set of closed loops as shown
in Fig.~\ref{fig:loop}(b).  To be concrete, we assign local graphs
according to the following rules: i) For an offdiagonal configuration,
at which the world line jumps, a horizontal graph
(\includegraphics*[width=1.0em]{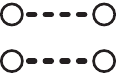}) is assigned with
probability 1. ii) For a diagonal configuration with antiparallel
spins ($\uparrow\downarrow$ or $\downarrow\uparrow$), a horizontal
graph is assigned with probability $\Delta_\tau/2$, otherwise a
vertical graph (\includegraphics*[width=1.0em]{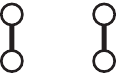}) is
assigned. iii) For a diagonal configuration with parallel spins
($\uparrow\uparrow$ or $\downarrow\downarrow$), a vertical graph is
assigned with probability 1.  This procedure is called {\em labeling}
or {\em breakup}.

Next, we flip each loop with probability 1/2.  By this {\em cluster
  flipping} procedure, spins on each loop are flipped
simultaneously from up to down or vice versa~[Fig.~\ref{fig:loop}(c)],
and as a result a new world-line configuration is
obtained~[Fig.~\ref{fig:loop}(d)].  It is straightforward to prove
that the labeling procedure together with the cluster flipping
procedure fulfills the balance condition as well as the ergodicity, so
the Monte Carlo average of any physical quantities is guaranteed to
converge to the correct value.  Furthermore, the extent of loops
corresponds directly to the antiferromagnetic spin correlation length
in space-time.  As a result, the correlations between successive
world-line configurations are removed almost
completely~\cite{Evertz2003,KawashimaH2004,Todo2012}, and the
autocorrelation time stays about few tens even for the largest systems
we simulated below.

One should note that the labeling probability of a horizontal graph to
the diagonal configuration is of ${\cal O}(\Delta_\tau)$ and thus the
total number of horizontal graphs also stays finite in the Trotter
limit.  In the practical implementation of the loop algorithm,
therefore, we can work directly in the imaginary time
continuum~\cite{BeardW1996}, which completely eliminates systematic
errors due to the imaginary time discretization.  To wrap up, one
Monte Carlo step in the continuous imaginary time loop algorithm is as
follows:
\begin{enumerate}
\item $\tau \leftarrow 0$ and $s \leftarrow (s_1,s_2,\cdots,s_L$) at
  $\tau=0$.
\item Draw a random number $r$ uniformly distributed in $(0,1]$, and
  $\tau' \leftarrow \tau - (2/L) \log r$.
\item If there are offdiagonal operators between $\tau$ and $\tau'$,
  assign a horizontal graph to them and update $s$ by flipping spins
  accordingly.
\item If $\tau' > \beta$, go to step 7.
\item Draw a random integral number $l$ uniformly distributed in
  $(0,L]$, and insert a horizontal graph between sites $l$ and
  $l+1$ at imaginary time $\tau'$, if $s_l \ne s_{l+1}$.
\item $\tau \leftarrow \tau'$ and goto step 2.
\item Identify clusters, flip them independently with probability 1/2,
  and update spin configurations at $\tau=0$ and series of offdiagonal
  operators accordingly.
\item Perform measurement of physical quantities.
\end{enumerate}
By inserting horizontal graphs in step 3 and 5, the space-time is
decomposed into many small fragments of vertical lines
[Fig.~\ref{fig:loop}(b)].  Although each loop is a sequence of such
fragments, it is more convenient to represent each loop by a rooted
tree of segments instead of a sequential list.  During step 2--6,
trees are merged to build up loops by using the union-find
algorithm~\cite{CormenLRS2001}.  It is proved that using two
techniques, {\em union-by-weight} and {\em path compression}, any
sequence of $m$ union and find operations on $n$ objects takes ${\cal
  O}(n \alpha(m,n))$ where $\alpha(m,n)$ is the inverse Ackerman
function that grows extremely slow with increasing $m$ and $n$.  In
any practical applications, $\alpha(m,n)$ is less than 5 and one may
regard it as a constant.

\section{Parallelization of Loop Algorithm}
\label{sec:Parallelization}

The number of world-line jumps as well as the number of horizontal
graphs are both proportional to $L \beta$, so is the total
amount of the memory required to store the world-line configuration
and the loop configuration.  The number of operations required for one
Monte Carlo sweep is also proportional to $L \beta$ as
discussed in the last section.  Since the autocorrelation time of the
present loop algorithm is so short [$\sim {\cal O}(10)$] that the best
parallelization strategy is running independent Markov chains with
using different random number sequences on every node, as long as the
memory requirement is not very strong.  If $L \beta \gtrsim
10^7$, however, the world-line configuration no longer fits in the
memory of one node, and parallelization of each Markov chain becomes
unavoidable.

\begin{figure}[tb]
\centering \includegraphics[width=2.8in]{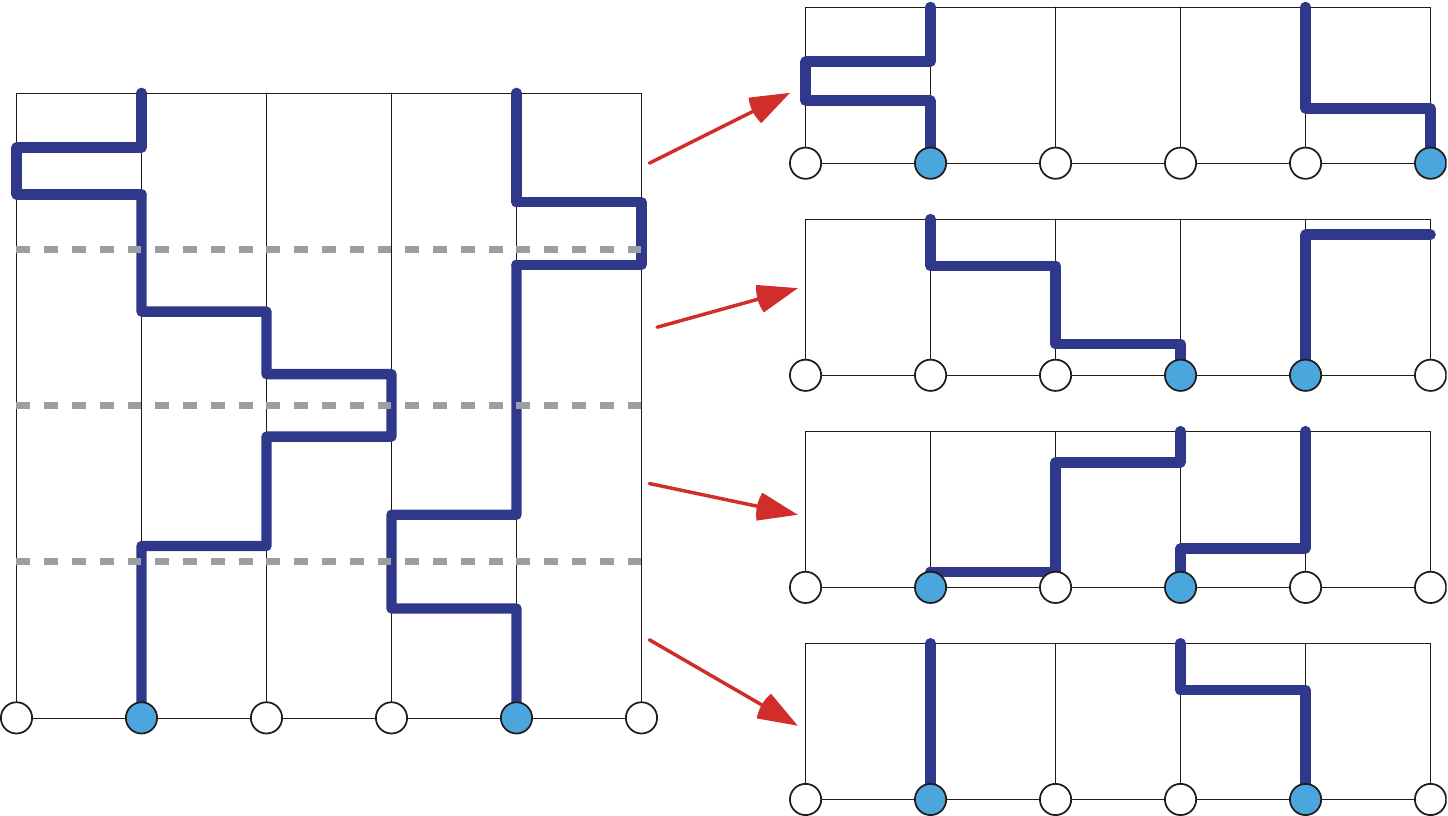}
\caption{An example of MPI domain decomposition of world-line
  configuration for $L=6$ and 4 nodes.  The horizontal and vertical
  axes are the spatial and imaginary time directions, respectively.
  The space-time is divided into slices of same thickness.  Each node
  stores the spin directions at $\tau = \beta p/N_{\rm p}$ (circles)
  together with space-time positions of world-line jumps (horizontal
  blue lines) in its own time window.}
\label{fig:ProcessDecomposition}
\end{figure}

\subsection{Basic strategy}
\label{sec:BasicStrategy}

In general, parallelization of the loop algorithm is far from trivial,
since a number of global objects, loops, are built up and flipped in
every Monte Carlo step.  The basic strategy for the nontrivial
parallelization has been given in Refs.~\cite{Todo2002} and
\cite{Todo2003}.  In order to parallelize the loop algorithm, first
we have to determine how the information of world-line configuration
is distributed to several nodes. In the continuous time loop algorithm
for one-dimensional quantum spin systems, it is most natural to divide
the (1+1)-dimensional space-time into $N_{\rm p}$ slices of thickness
$\beta/N_{\rm p}$ in the imaginary time direction, where $N_{\rm p}$
is the number of nodes.  Each node stores only the spin directions at
imaginary time $\beta p/N_{\rm p}$, where $p$ is the processor index
($p=0,1,\cdots,N_{\rm p}-1$) together with space-time positions of
world-line jumps in its own imaginary time
window~(Fig.~\ref{fig:ProcessDecomposition}).
The same imaginary-time decomposition scheme can be used for higher-dimensional systems as well.
One of the main
advantages of this domain decomposition scheme is that the {\em
  thickness} of each slice can be chosen as the same regardless of the
number of nodes or the network topology, as the system is continuous
in the imaginary time direction.  Another advantage of the present
decomposition is that the algorithm does not depend on the
dimensionality of the system.

\begin{figure}[tb]
\centering \includegraphics[width=4in]{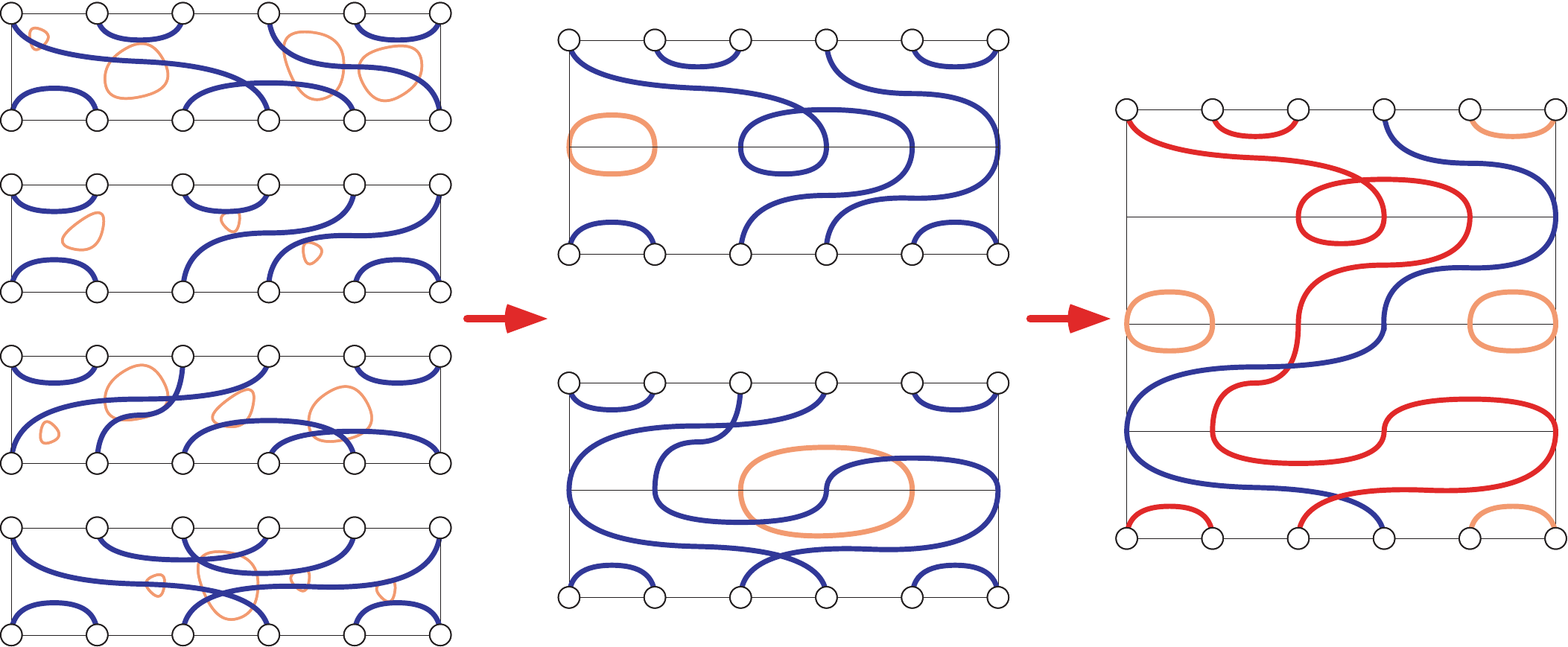}
\caption{An example of the binary-tree global union-find procedure for
  $L=6$ and 4 nodes.  Orange lines denote locally-closed loops, which
  can be ignored in the succeeding stages.  In this case, one finally
  finds two global loops (red and blue loops in the rightmost figure)
  wrap around the whole system in the imaginary time direction.}
\label{fig:BinaryTree}
\end{figure}

By adopting the above domain decomposition scheme for the world-line
configuration, the labeling procedure can be trivially parallelized
and executed independently, since it requires local information only,
i.e., the relative direction of two neighboring spins at the same
imaginary time.  No internode communication is required at this stage.
On the other hand, identifying loops is nontrivial, since loops are
global objects.  We perform the global union-find procedure in the
following two steps: (i) each node identifies loops in its own
imaginary time window.  In this step, a number of loops are closed in
the time window.  However, there remain $L$ unclosed loops, since they
cross the imaginary-time boundaries between the nodes.  This step
requires no internode communication as the same as the labeling stage.
Then, (ii) these unclosed loops are merged gradually by using a
binary-tree algorithm as shown schematically in
Fig.~\ref{fig:BinaryTree}.  Finally, the information on global loops
is distributed back to all the nodes to determine the next world-line
configuration.  The number of operations required to the labeling
process and the step (i) of the union-find process is proportional to
$L\beta$ and these operations are ideally parallelized.  On
the other hand, the step (ii) of the union-find process can not be
executed independently.  The number of operations on the master node is
proportional to $L\log_2 N_{\rm p}$.  Thus, the theoretical
efficiency, $P(N_{\rm p})$, of the present parallelized loop algorithm
is evaluated as
\begin{align}
  \label{eqn:scaling}
  P(N_{\rm p}) = \frac{\beta L}{N_{\rm p}} \Big/ \Big[ \frac{\beta
      L}{N_{\rm p}} + c L \log_2 N_{\rm p} \Big] \simeq 1 - c
  \frac{N_{\rm p}}{\beta} \log_2 N_{\rm p}
\end{align}
for $\beta \gg N_{\rm p}$, where $c$ is a system-dependent constant.

Based on the above strategy, the parallel loop algorithm was first
implemented using flat MPI, and it was confirmed that the code scales
fairy well up to about $10^3$ nodes~\cite{Todo2002,Todo2003}.  In the
environment with more than $10^4$ nodes ($10^5$ cores), however, it
turns out that the overhead due to parallelization becomes
non-negligible, and the simulation code does not scale any more.  To
overcome the difficulty, we have developed and implemented the
following parallelization techniques.

\subsection{Asynchronous wait-free union-find algorithm}
\label{sec:AsynchronousUnionFind}

\begin{figure}[tb]
  \centerline{\includegraphics[width=6.5in]{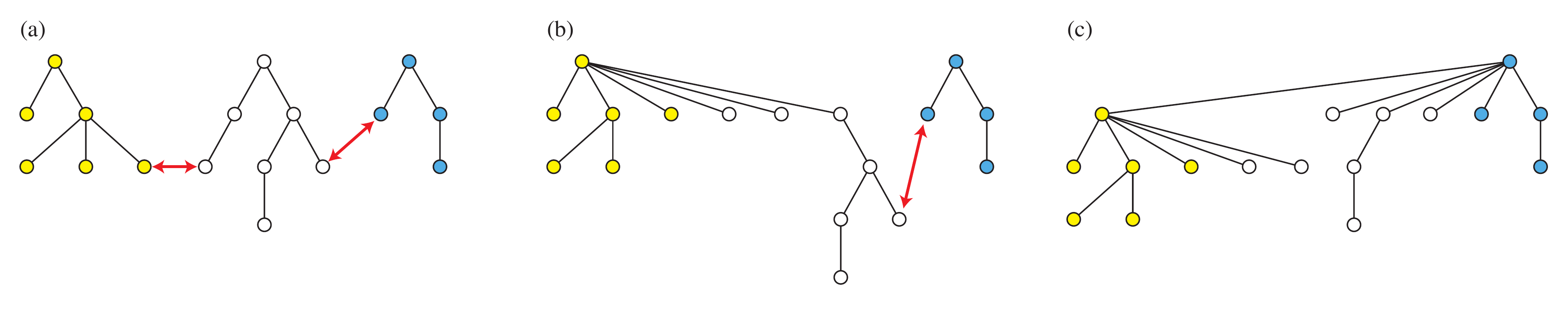}}
  \caption{Example of a union-find procedure in which simultaneous update of trees can break the tree structures.
    The parent pointer of root vertex in the white tree can be rewritten simultaneously by two different threads.}
\label{fig:union-find}
\end{figure}

The first improvement is the introduction of hybrid parallelization
using OpenMP together with MPI.  First of all, by introducing the
MPI-OpenMP hybrid parallelization, the amount of memory used by the
MPI library is reduced greatly compared with the case of using flat
MPI parallelization.  Second, introducing the parallelization with
respect to another axis, the efficiency of parallelization can also be
improved.  In addition to the MPI process parallelization based on the
domain decomposition in the imaginary time direction, we introduce
OpenMP thread parallelization with respect to the real-space
direction; the one-dimensional chain lattice is decomposed into
domains with equal number of bonds (edges).  Each thread maintains the
list of operators on the bonds in its own domain, while the tree
structure representing loops is shared globally by all threads in an
MPI process.  Since the union-find operations performed in each thread
may alter the whole structure of trees, exclusive access control with
high granularity must be introduced.

A union-find procedure of connecting two vertices A and B consists of
the following steps:
\begin{enumerate}
\item Find the root vertex of A by tracking parent of vertices.
\item Find the root vertex of B by tracking parent of vertices.
\item Return if A and B belong to the same tree.
\item Compare the weight (number of vertices in each tree) of the root
  vertices, and choose a new root vertex.
\item Update the weight of the new root vertex.
\item Update the parent pointer of the other (new leaf) vertex.
\item Compress the path from vertices A and B to the new root vertex by
  rewriting the parent pointers of the vertices on the paths.
\end{enumerate}
First, it should be noticed that in the union-find algorithm, once a
vertex becomes a leaf vertex of a tree, then it never becomes back to
a root vertex.  Therefore, {\em loops} will never be created even if
multiple threads try to alter the same tree simultaneously without
access control.  This guarantees that step 1--4 and 7 are already {\em
  thread-safe}, and we don't need to modify these parts of the
algorithm.  On the other hand, we need to introduce some {\em locking}
mechanism for step 5 and 6 to avoid breaking the tree structure by
simultaneous updates of the parent pointer of the same root vertex by
multiple threads (Fig.~\ref{fig:union-find}).  The lock should be done
vertex-wise so that the operation on different trees can be performed
concurrently.  At the same time, it should not require any extra
memory, since the number of vertices are of ${\cal O}(L\beta)$.  To
this end, we have implemented the locking mechanism of each vertex by
the inline assembler using the {\em compare-and-swap} atomic
instruction ({\tt cmpxchgl} in Intel64 architecture and {\tt cas} in
SPARC architecture).  Thus, in the present thread-safe version of
union-find algorithm, we modify step 5 and 6 as
\begin{enumerate}
\item[4')] Try to lock the two root vertices.  If one or both lock trials
  fail, repeat from step 1.
\item[5)\ ] Update the weight of the new root vertex.
\item[6)\ ] Update the parent pointer of the other (new leaf) vertex.
\item[6')] Release the lock of the vertices.
\end{enumerate}
This algorithm is {\em thread-safe} and {\em wait-free}.  In practice,
we implement this algorithm with no extra memory space by reusing the
parent pointer (and weight) of each vertex as a lock object.
Furthermore, the release of locks (in step 6') will be done
automatically by setting a new weight or the parent pointer in step 5
and 6.  By this way, we can minimize the time window in which one or
more vertices are in the {\em locked} state.

\subsection{Butterfly-type global union-find algorithm}
\label{sec:ButterflyUnionFind}

\begin{figure}[tb]
\centering \includegraphics[width=\textwidth]{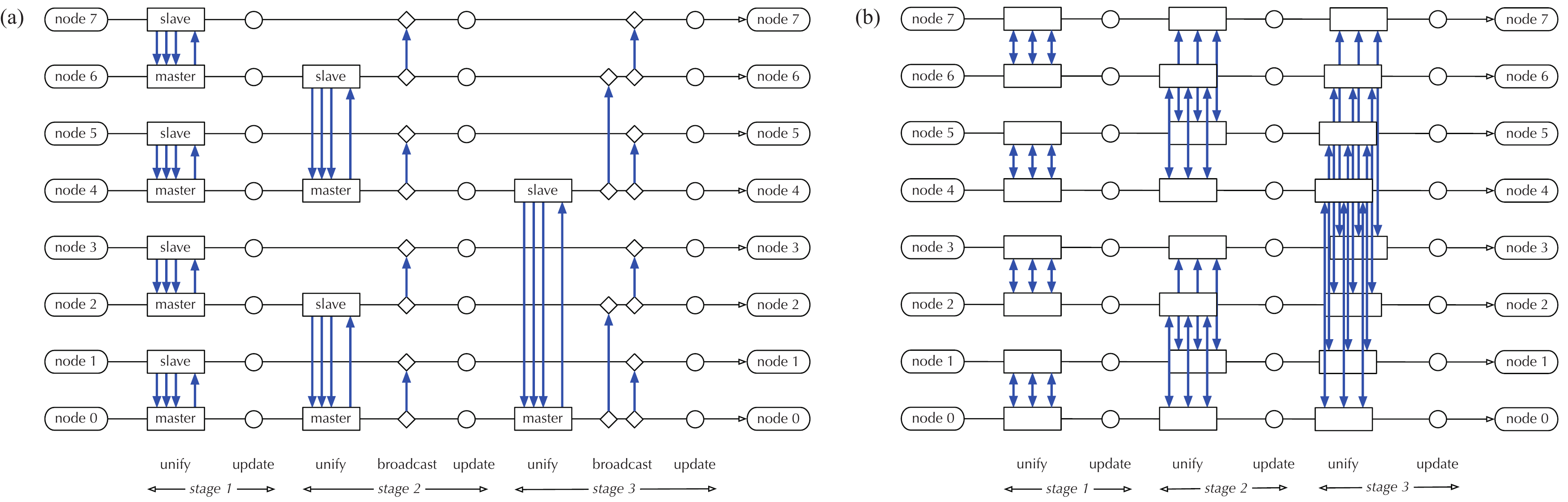}
\caption{Patterns of communication in global union-find procedure
  using (a) binary-tree type and (b) butterfly type algorithms.  In
  the former algorithm, the results of union-find at each stage have
  to be broadcasted to all the descendent nodes, which dominates the
  overall performance as the number of stages increases.  On the other
  hand, in the latter algorithm combined with the majority-vote trick,
  the cost for each stage stays constant on hyper-torus networks.}
\label{fig:CommunicationPattern}
\end{figure}

The binary-tree global union-find algorithm described in
Sec.~\ref{sec:BasicStrategy} works fairy well up to $N_{\rm p} \simeq
10^3$~\cite{Todo2002,Todo2003,TodoK2001}.  However, as the number of
nodes is increased further, the overhead due to parallelization
becomes non-negligible.  The most dominant overhead is the cost for
broadcasting the results of union-find procedure at each stage
[Fig.~\ref{fig:CommunicationPattern}(a)].  At each intermediate stage,
$2L$ loop fragments are unified and as a result $L$ open loop
fragments and a number of closed loops are obtained
(Fig.~\ref{fig:BinaryTree}).  For the closed loops, we have to
determine whether each of them will be flipped or not to generate the
next world-line configuration, and broadcast the decision immediately
to all the descendent nodes, since there is no memory space to store
all the decisions until the whole union-find procedure is completed.
Also, for the open loop fragments, we have to maintain the
correspondence table between the original loop fragments to the new
ones, and broadcast it to all the descendent nodes.  For example, for
$N_{\rm p} = 24,576$, the total number of stages becomes 14 or 15, and
at each stage we have to perform broadcast of the same order of {\rm
  depth}.

In order to avoid the overhead of broadcast, we develop a
butterfly-type global union-find algorithm.  In this algorithm, the
pattern of communication is modified as shown in
Fig.~\ref{fig:CommunicationPattern}(b), where instead of sending the
loop fragment information from a slave to a master at each stage, two
(or more) nodes exchange the loop fragment information with each
other, then each of them performs the same union-find operations.  By
this way, broadcasting of the correspondence table is completely
eliminated.

Note that in this butterfly-type algorithm, the union-find of loop
fragments will be done {\em out of order} on each node.
Accordingly, the tree structure built up on each node will also look
completely different with each other, though final grouping of
vertices (represented by trees) does not depend on the order of
union-find operations and should be identical after the final stage
has been completed.  This fact makes it difficult for us to make
globally consistent decisions on the loop flip, since the random
number used for decisions generally depends on the execution order.

This difficulty is solved by introducing another technique, {\em
  majority-vote trick}.  Instead of postponing the decision until loop
fragments form a closed loop, we vote positive or negative on all loop
fragments on each node in advance.  The vote for each loop fragment is
exchanged together with the other loop fragment information between
nodes, and the number of positive and negative votes is counted in
parallel with the union-find operation to make a final decision on the
loop flip.  Since the result of vote counting does not depend on its
order, the decision for each loop will be globally consistent
irrespective of the order of union-find operations.  In practice,
instead of counting positive and negative votes, we use XOR
(exclusive-or) operations to tally the votes made for each loop in
order to avoid ending in a draw accidentally.

\subsection{Optimized process mapping on finite-dimensional torus}
\label{sec:ProcessMapping}

The butterfly-type global union-find algorithm works efficiently on a
network with hypercube topology of dimension $\log_2 N_{\rm p}$.  On a lower-dimensional torus
network, however, butterfly-type communication between distant nodes causes a network
congestion in general, and the network transfer performance is often
spoiled substantially.  Moreover, if the linear extent of the torus
network is not a power of 2, communications along different axes of
the torus also interfere with each other.  For example, the Tofu
interconnect~\cite{AjimaSS2009,Toyoshima2010} of the K computer
logically provides a three-dimensional torus to users, and if a user
wants to run a job by using the whole K computer, $48 \times 54 \times
32$ is the only possible shape of the three-dimensional torus, which
is factored into $(2^4 \times 3) \times (2 \times 3^3) \times 2^5$.
In order to realize communication that is free from interference
between different axes, one has to combine a pair-wise communication
with three-point communication.

We have extended the global union-find algorithm, so that it works on
generic virtual toruses of any dimensions, i.e., $\ell_1 \times \ell_2
\times \cdots \times \ell_d$ with any set of positive integers
$(\ell_1,\ell_2,\cdots,\ell_d)$.  In $k$-th dimension of the virtual
torus, which forms a set of periodic chains of length $\ell_k$, the
union-find operations are repeated $\lfloor \ell_k/2 \rfloor$ times (steps) by
using only the two-sided nearest neighbor communication on that chain.
In the above example, one can define the virtual torus as
$(\ell_1,\ell_2,\ell_3)=(48,54,32)$, in which we have no congestion
but the total number of steps is relatively large ($24+27+16=67$).
Another possibility is
$(\ell_1,\ell_2,\cdots,\ell_{14})=((2,2,2,2,3),(2,3,3,3),(2,2,2,2,2))$, by
which the total number of steps is minimized (14).
There are many
other possibilities between the extrema.  There is a trade-off
between the number of steps and severity of network congestion, and
the optimal choice will depend on the system.  
Note that in the second example of the virtual torus, we introduce
nested parentheses to emphasize that the product of lengths in each
group is consistent with that of the original torus, and thus the
inference between different axes of the original torus is avoided.
  
\section{Performance Analysis}
\label{sec:PerformanceAnalysis}

Based on the techniques presented in the last section, we have
implemented the hybrid parallel version of the continuous time loop
algorithm (ALPS/looper version 4~\cite{LOOPERweb}) from the scratch by
using the C++ programming language.  The simulation code has been
tested on 24,576 nodes (196,608 cores) of the K computer.

\subsection{OpenMP thread parallelization}

\begin{figure}[tb]
  \centering \includegraphics[height=2.7in]{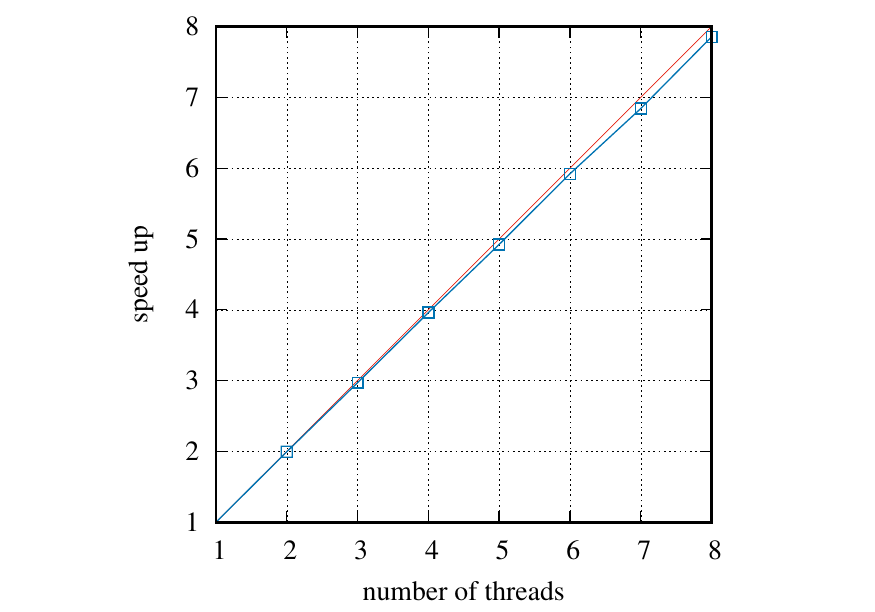}
\caption{Strong scaling test of the asynchronous wait-free union-find
  algorithm in the Swendsen-Wang cluster
  algorithm for two-dimensional square lattice
  Ising model with $L=8,192$ and $K=K_{\rm c} = 0.440686\cdots$.
  The red line indicates an ideal speed-up.}
\label{fig:sw-strong-scaling}
\end{figure}

First, we test the efficiency of the asynchronous wait-free union-find
algorithm, introduced in Sec.~\ref{sec:AsynchronousUnionFind}.
As a benchmark test, we adopt the Swendsen-Wang cluster algorithm~\cite{SwendsenW1987} for the two-dimensional square lattice Ising model.
This algorithm is much simpler than the loop cluster algorithm, and thus more suitable for evaluating the performance the union-find algorithm directly.
In the Ising model, we have
classical spins $\sigma = \pm 1$ aligned to form an $L \times L$
two-dimensional array.  In the Swendsen-Wang algorithm, the nearest
neighbor sites are connected with probability $p=1-e^{-2K}$ if the
spins on these sites have the same value.  Then, the spins on each
cluster are simultaneously flipped with probability 1/2.  We choose
the parameter $K$ as $K_{\rm c} = \ln(1+\sqrt{2})/2 = 0.440686\cdots$,
the critical point of this model, where the average cluster size,
which corresponds to the magnetic susceptibility diverges as $L^{7/4}$
with increasing the system size $L$~\cite{Goldenfeld1992}.

A benchmark test has been executed on a single node of the K computer.
In Fig.~\ref{fig:sw-strong-scaling}, we show the result of the strong scaling test for $L=8,192$.
We measure the execution time of the union-find operations during 128 Monte Carlo steps after discarding first 128 steps as burn-in time.
The number of threads is 1, 2, 3, 4, 6, and 8.
It takes about 0.097 second per Monte Carlo step in the single thread case.
As one can see clearly in Fig.~\ref{fig:sw-strong-scaling}, almost perfect strong scaling has been achieved up to 8 threads.
The parallelization efficiency is about 98.2\% if one compares the 8-thread performance with the single thread case.

\subsection{Network performance}

\begin{table}[tb]
\caption{Network transfer performance on 12,288 nodes of the K
  computer.  The shape of three-dimensional torus is $32 \times 12
  \times 32$ and the shape of virtual lattice (transfer pattern) is
  chosen as $((2,2,2,2,2),(2,2,3),(2,2,2,2,2))$.}
\label{tbl:network}
\centering
\begin{tabular}{ccccc}
\hline
stage & $\ell_k$ & transfer rate [GB/sec] & multiplicity & efficiency [\%] \\
\hline
\hline
1 & 2 & 4.44 & 1 & 88.8 \\
2 & 2 & 2.26 & 2 & 90.4 \\
3 & 2 & 1.13 & 4 & 90.4 \\
4 & 2 & 0.56 & 8 & 89.6 \\
5 & 2 & 0.56 & 8 & 89.6 \\
6 & 2 & 4.44 & 1 & 88.8 \\
7 & 2 & 2.26 & 2 & 90.4 \\
8 & 3 & 1.50 & 4 & 120.0 \, \\
9 & 2 & 4.32 & 1 & 86.4 \\
10 & 2 & 2.26 & 2 & 90.4 \\
11 & 2 & 1.13 & 4 & 90.4 \\
12 & 2 & 0.56 & 8 & 89.6 \\
13 & 2 & 0.56 & 8 & 90.6 \\
\hline
\end{tabular}
\end{table}

We demonstrate the effectiveness of the process mapping strategy
introduced in Sec.~\ref{sec:ProcessMapping} by using 12,288 nodes of
the K computer.  The shape of three-dimensional torus is $32 \times 12
\times 32$.  We choose a 13-dimensional virtual lattice whose shape is
given by $((2,2,2,2,2),(2,2,3),(2,2,2,2,2))$, so that
there occurs no interference between different axes.  In the benchmark
program, a continuous data of 10~MB is transferred in each stage in
both directions between nodes.  In Table.~\ref{tbl:network}, we
summarize the network transfer performance at each stage.  At stage 1,
6, 9, the data transfer is performed between adjacent nodes on the
three-dimensional torus.  In these cases the transfer rate is
4.32--4.44~GB/s which is about 86.4--88.8\% of the theoretical peak bandwidth
5~GB/s.  At stage 2, 7, 10, each link between the nodes is shared by
two communications.  Taking into account that the multiplicity of the
link is 2, the efficiency is evaluated as 90.4\%.
The multiplicity of each stage is given by the product of lengths ($\ell_k$) of earlier stages in the current axis. In the final stage of each axis, if $\ell_k=2$, the multiplicity becomes half, since the other side of the periodic chain can be utilized simultaneously. For example, we estimate the multiplicity of stage 4 (13) is $2 \times 2 \times 2 = 8$ ($2 \times 2 \times 2 \times 2 \,/\, 2 = 8$).
Remarkably, the
efficiency is not degraded even for the case with 8-fold multiplicity
(stage 4, 5, 12, 13).  The efficiency is as high as the case of
adjacent communication, and there is absolutely no signs of network
congestion.

We should point out that at stage 8, which has $\ell_k=3$ and 4-fold
multiplicity according to the above rule, shows performance much
higher than the theoretical peak. This unforeseen performance may owe to the
physical structure of the Tofu
network, i.e., 6-dimensional mesh/torus of $32^3 \times 2 \times 3 \times 2$~\cite{AjimaSS2009,Toyoshima2010}. Indeed, by assuming 90~\% of
the theoretical peak, we can estimate the effective multiplicity as
\begin{equation}
  5 \times 0.9 / 1.50 \simeq 3,
\end{equation}
which manifests the existence of extra links hidden from the
three-dimensional logical view of the torus.

\subsection{Weak scaling property}

\begin{figure}[tb]
\centering
\includegraphics[height=2.7in]{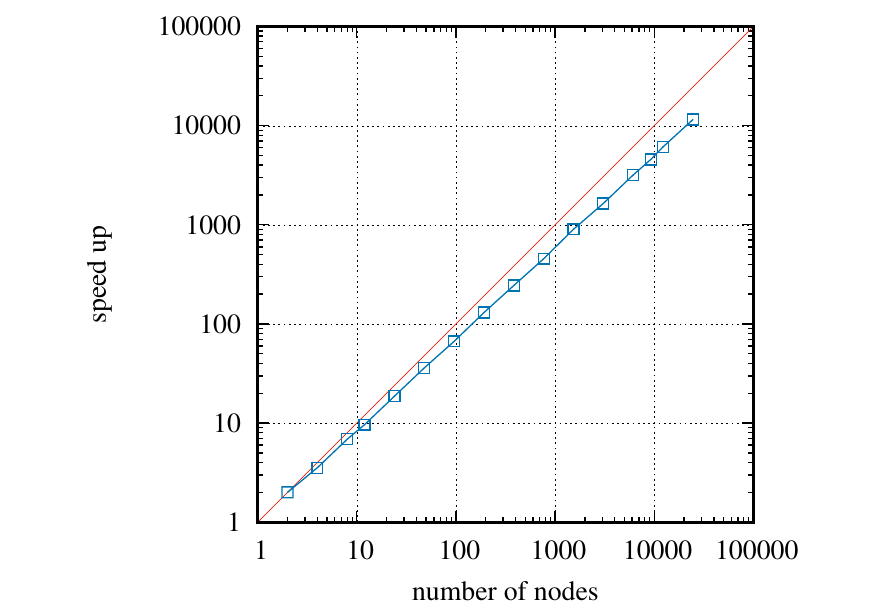}
\caption{Weak scaling
  results for $N_{\rm p}=2$, 4, 8, 12, 24, 48, 96, 192, 384, 768,
  1,536, 3,072, 6,144, 9,216, 12,288, and 24,576.  The parallelization
  efficiency for $N_{\rm p} = 24,576$ is about 46.9\%.  The red line
  denotes the ideal speed-up.}
\label{fig:loop-weak-scaling}
\end{figure}

The efficiency of the present OpenMP-MPI hybrid parallelized
continuous time loop algorithm is demonstrated by the weak scaling
test up to 24,576 nodes (196,608 cores) of the K computer.  In this
weak scaling test, the system size is fixed to $L=2,621,440$ and the
inverse temperature is increased in proportional to the number of
nodes as $\beta = 12.642 N_{\rm p}$.  The number of threads in each
process is 8, and the number of nodes, which is the same as the number
of processes, is tested starting from $N_{\rm p}=2$ up to 24,576.  The
overall speed-up as a function of $N_{\rm p}$ is shown in
Fig.~\ref{fig:loop-weak-scaling}, where the speed-up at $N_{\rm p}=2$
is normalized to 2.  With $N_{\rm p} = 24,576$, we have achieved
speed-up by a factor of $1.1 \times 10^4$.  The parallelization
efficiency is thus about 46.9\%.

\begin{figure}[tb]
\centering
\includegraphics[height=2.7in]{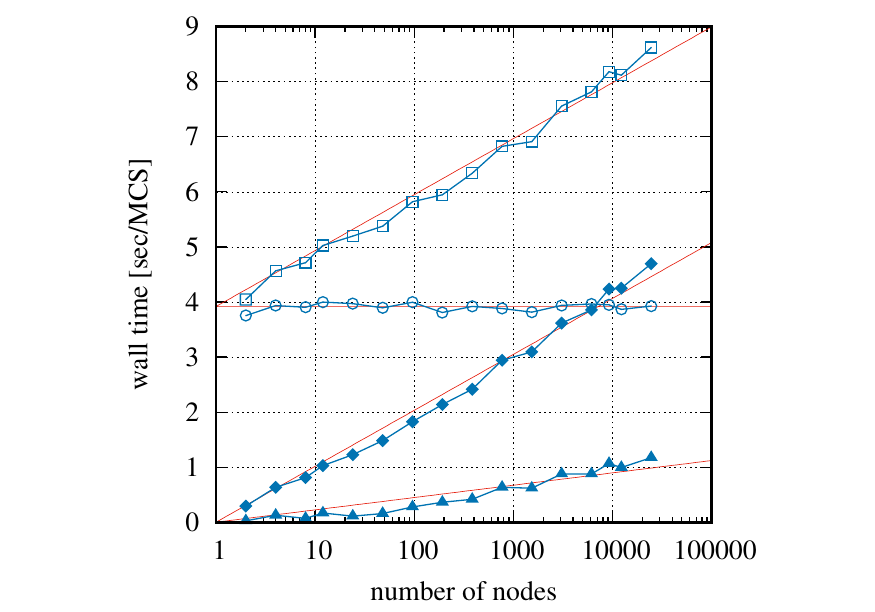}
\caption{$N_{\rm p}$-dependence of the cost of each section: whole
  Monte Carlo step (squares), labeling and local union-find (circles),
  global union-find including communication (filled diamonds),
  communication (filled triangles).  The cost for global union-find as
  well as that for communication is roughly proportional to $\log_2
  N_{\rm p}$.}
\label{fig:loop-section}
\end{figure}

The decline of the efficiency is mainly due to the overhead of global
union-find operations.  In Fig.~\ref{fig:loop-section}, we show the
$N_{\rm p}$-dependence of the cost of each section.  One sees that the
cost of global union-find operation as well as that of the
communication grow in proportional to $\log_2 N_{\rm p}$, while the cost
of labeling and the local union-find operations stays constant
irrespective of the number of nodes.  This behavior is consistent with
the theoretical estimation given in the last section
[Eq.~(\ref{eqn:scaling})].

\begin{table}[tb]
\caption{Cost distribution in one Monte Carlo step for 12,288 and 24,576 nodes on the K computer.}
\label{tbl:costdistribution}
\centering
\begin{tabular}{lcc}
\hline
& 12,288 nodes [sec] & 24,576 nodes [sec] \\
\hline
\hline
1) random number generation in exponential distribution & 0.53 & 0.54 \\ % 3+4
2) insert/remove operators and local union-find operations & 1.77 & 1.83 \\ % 5+6+7+8
3) assignment of loop IDs & 0.33 & 0.33 \\ % 10
4) accumulation physical properties of loops & 0.91 & 0.91 \\ % 11+12+13
5) global union-find operations (except for communication) & 3.26 & 3.52 \\ % 14-29-30
6) pair-wise communication & 0.91 & 1.09 \\ % 30
7) three-point communication & 0.08 & 0.09 \\ % 29
8) update of spins and operators & 0.32 & 0.32 \\ % 15
\hline
5+6+7 & 4.25 & 4.69 \\ % 14
total & 8.12 & 8.62 \\ % 2
\hline
\end{tabular}
\end{table}

The detailed comparison of the cost between the cases with $N_{\rm
  p}=12,288$ and 24,576 is given in Table~\ref{tbl:costdistribution}.
For $N_{\rm p}=12,288$, we choose the shape of three-dimensional torus as $16\times 24 \times 32$ and that of virtual lattice as $((2,2,2,2),(2,2,2,3),(2,2,2,2,2))$. For $N_{\rm p}=24,576$ the length along the first axis of the torus is doubled and the shape of virtual lattice is chosen as $((2,2,2,2,2),(2,2,2,3),(2,2,2,2,2))$.
We attribute the slight growth in section 2 to fluctuations between
nodes due to some system noise.  The growth of cost in section 6
(pair-wise communication), which is the summation of the communication cost of stages with $\ell_k=2$, should be explained based on the network
performance analysis in the previous subsection.  Taking into account
that the length along the first axis of the torus is doubled from 16 to
32, the growth of the cost is estimated as
\begin{equation}
  \frac{(1+2+4+8+8)+(1+2+4)+(1+2+4+8+8)}{(1+2+4+4)+(1+2+4)+(1+2+4+8+8)} \simeq 1.29,
\end{equation}
which is consistent with the experiments, $1.09/0.91 \simeq 1.20$.
Similarly, the growth of cost in section 5 [global union-find (except
  for communication)] ($3.52/3.26 \simeq 1.08$) can be explained by
the increase of the number of stages ($14/13\simeq 1.08$).

\subsection{Overall performance}

With 24,576 nodes of the K computer, we simulate a chain with
2,621,440 spins at inverse temperature 310,690.  The space-time volume
is thus $L \beta \simeq 8.14 \times 10^{11}$.  The average
number of operators (or horizontal graphs) is about $5.64 \times
10^{11}$.  One Monte Carlo step takes about 8.62 seconds.  In each
Monte Carlo steps, we generate a graph of $1.13 \times 10^{12}$
vertices (world-line fragments) and $1.13 \times 10^{12}$ edges, and
identify clusters by performing union-find operations on such a huge
graph, resulting $3.24 \times 10^{11}$ clusters in average.  Note that
size of the largest cluster is of the same order as the space-time
volume $\sim 10^{12}$ as the ground state of the present $S=1/2$
antiferromagnetic Heisenberg chain is critical and thus the
correlation length is infinite in the thermodynamic limit.

Since the most time consuming part in the present algorithm is the
union-find operations on the tree structure, the performance index
concerning floating-point operations is not impressive.  The overall
performance using 24,576 nodes is 7.63~TFLOPS (tera floating-point
operations per second) and 0.164~PIPS (peta instructions per second), which are 0.243\% and 10.4\% of
the theoretical peak performance, 3.15 PFLOPS ($=10^3$~TFLOPS) and 1.57~PIPS, respectively.

\begin{table}[tbp]
\caption{Ground-state energy density $E/L$, staggered susceptibility $\chi_{\rm s}$, spatial correlation length $\xi$, first excited gap $\Delta$ of $S=1$, 2, 3, 4 antiferromagnetic Heisenberg chains estimated by various numerical methods: MCPM (Monte Carlo Power Method), RSRG (Real Space Renormalization Group), QMC (Quantum Monte Carlo), DMRG (Density Matrix Renormalization Group), ND (Numerical Diagonalization).}
\label{tbl:gap}
\centering
\begin{tabular}{clllll}
\hline
$S$ & \multicolumn{1}{c}{$E/L$} & \multicolumn{1}{c}{$\chi_{\rm s}$} & \multicolumn{1}{c}{$\xi$} & \multicolumn{1}{c}{$\Delta$} & \multicolumn{1}{c}{method} \\
\hline
\hline
1 & -1.4015(5) & & & 0.41 & MCPM~\cite{NightingaleB1986} \\
&   -1.449724 & & & 0.368166 & RSRG~\cite{PanC1987} \\
&   -1.4021(5) & & & 0.4097(5) & RSRG~\cite{LinP1988} \\
&              & & 6.25 & 0.425 & QMC~\cite{Nomura1989} \\
&   -1.401484038971(4) & & 6.03(2) & 0.41050(2) & DMRG~\cite{WhiteH1993} \\
&   -1.401485(2) & & 6.2 & 0.41049(2) & ND~\cite{GolinelliJL1994} \\
&   -1.401481(4) & 18.4048(7) & 6.0153(3) & 0.41048(6) & QMC \cite{TodoK2001} \\
& & & & 0.41047777 & ND (lower bound) \cite{NakanoT2009} \\
& & & & 0.41048023 & ND (upper bound) \cite{NakanoT2009} \\
\hline
2 & & & 80 & 0.08 & QMC~\cite{DeiszJC1993} \\
& & & 80 & 0.02 & QMC~\cite{HatanoS1993} \\
& -4.7608 & & 33 & 0.02 & DMRG~\cite{QinNS1995} \\
& -4.7545 & & & 0.05 & QMC~\cite{Sun1995} \\
& -4.761248(1) & & 49(1) & 0.085(5) & DMRG~\cite{SchollwoeckJ1995} \\
&   -4.76125(5) & & & 0.082(3) & DMRG~\cite{QinLY1997} \\
&   -4.761244(1) & & 54.3(2) & 0.085(1) & DMRG~\cite{QinWY1997} \\
& & $1.16(1) \times 10^3$ & 50(1) & 0.090(5) & QMC~\cite{KimGWB1997} \\
& & & & 0.0907(2) & DMRG~\cite{Schollwoeck1999} \\
& & & & 0.0876(13) & DMRG \cite{WangQY1999} \\
& -4.761249(6) & $1.1640(2) \times 10^3$ & 49.49(1) & 0.08917(4) & QMC \cite{TodoK2001} \\
& & & & 0.0878 & ND (lower bound) \cite{NakanoT2009} \\
& & & & 0.0896 & ND (upper bound) \cite{NakanoT2009} \\
& & & 49.6(1) & 0.0891623(9) & DMRG~\cite{UedaK2011} \\
& & & & 0.0884 & ND (lower bound) \cite{NakanoS2018} \\
& & & & 0.0896 & ND (upper bound) \cite{NakanoS2018} \\
\hline
3 & -10.1239(1) & $1.580(3) \times 10^5$ & 637(1) & 0.01002(3) & QMC \cite{TodoK2001}\\
& & & & 0.0082 & ND (lower bound) \cite{NakanoT2009} \\
& & & & 0.0102 & ND (upper bound) \cite{NakanoT2009} \\
\hline
4 & & & & $6.3 \times 10^{-4}$ & ND (lower bound) \cite{NakanoT2009} \\
& & & & $8.1 \times 10^{-4}$ & ND (upper bound) \cite{NakanoT2009} \\
& -17.480(7) & $3.49(4) \times 10^7$ & $1.040(7) \times 10^4$ & $7.99(5) \times 10^{-4}$ & QMC (present work) \\
\hline
\end{tabular}
\end{table}

\section{Estimation of the Haldane Gap of $S=4$ Chain}
\label{sec:SimulationResults}

The ground state of the $S=1/2$ antiferromagnetic Heisenberg chain
[Eq.~(\ref{eqn:Hamiltonian})] is known to be critical, where the
antiferromagnetic correlation function decays algebraically as the
distance increases, and above the ground state there are gapless
excitations~\cite{desCloizauxP1962}.  On the other hand, in the
classical limit ($S \rightarrow \infty$), each spin behaves as a
classical vector, and the ground state is the long-range ordered N\'eel
state.  It again has gapless spin-wave excitations.

In 1982, Haldane~\cite{Haldane1982} made a striking conjecture that
the antiferromagnetic Heisenberg chain with integer $S$
($S=1,2,\cdots$) has a {\em finite} excitation gap $\Delta(S)$ above
its unique ground state, and the antiferromagnetic spin correlation
along the chain decays {\em exponentially} with a finite correlation
length $\xi(S)$.  For the $S=1$, 2, and 3, the finiteness of the first
excitation gap as well as the correlation length has already been
established numerically as listed in Table~\ref{tbl:gap}.  Simulations of higher-spin
systems, however, are much harder since the excitation gap (the
correlation length) becomes exponentially small (large) as $S$
increases~\cite{Haldane1982}.  The main difficulty for measuring such
a small gap is that the system behaves as if gapless (critical) so
long as the temperature is not low enough compared to the gap.
Similarly, the system size used in the simulation should be much larger
than the correlation length to detect a finite correlation length.
Thus we need to simulate extremely large systems at extremely low
temperatures in order to access the low-energy properties of the
higher-spin systems, and precise estimation of $\Delta(S)$ and
$\xi(S)$ for $S=4$, 5, $\cdots$ is still being one of the most
challenging problems in the computational statistical and
condensed-matter physics.

Let us consider the antiferromagnetic Heisenberg
chain~(\ref{eqn:Hamiltonian}) with $S=4$.  In order to apply the
continuous time loop algorithm to the this system, first we represent
the $S=4$ system by an equivalent $S=1/2$ system; each $S=4$ spin is
represented as a composition of 8 $S=1/2$ spins (subspins), and
simultaneously each bond is transformed into 64 bonds of the same
strength connecting subspins~\cite{HaradaTK1998,TodoK2001}.  The
largest system we simulate is $L=73,728$ ($=2^{13} \times 3^2$), which is converted in
advance into an equivalent $S=1/2$ system of $8 \times 73,728$ sites
and $64 \times 73,728$ bonds.  The inverse temperature is chosen as
$\beta = L / (2S) = 9216$.

\begin{figure}[tb]
\centering
\includegraphics[height=2.7in]{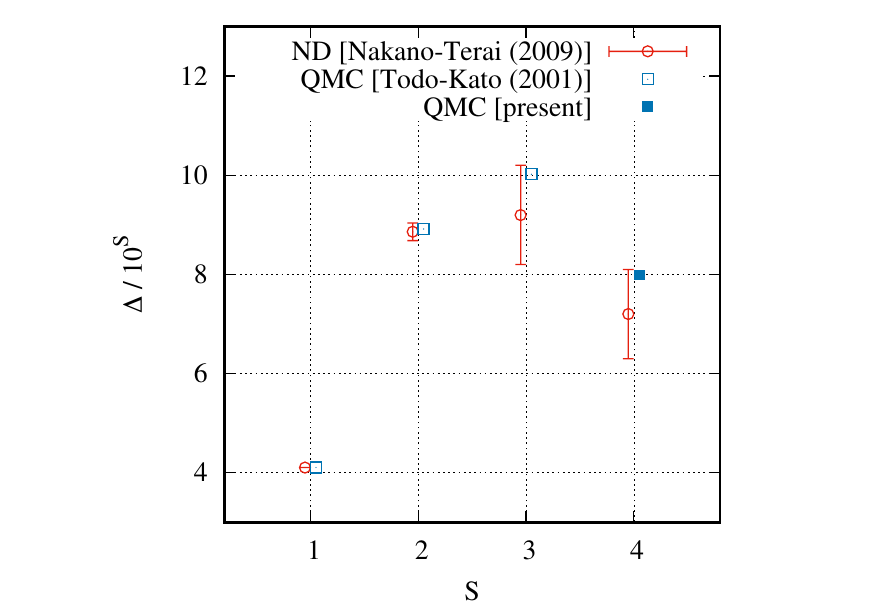}
\caption{$S$-dependence of the first excitation gap of the spin-$S$
  antiferromagnetic Heisenberg models calculated by the quantum Monte
  Carlo method (blue symbols)~\cite{TodoK2001}.  Note that $\Delta
  \times 10^S$ is plotted instead of $\Delta$ itself, since it
  decreases exponentially as $S$ increases.  The error bar of each
  data point is smaller than the symbol size.  A lower and upper bounds
  estimated by the numerical diagonalization (ND) are also indicated by red
  lines~\cite{NakanoT2009}.}
\label{fig:gap}
\end{figure}

The simulation is performed by using 2,048 nodes.  About 10~GB of
memory is required per node ($\sim$ 20~TB in total) in order to store
the world-line and loop configurations.  The correlation
length and the magnitude of the gap are calculated by means of the higher-order version of the
moment estimators~\cite{TodoK2001,SuwaT2015}.  The measurement is performed
during 8,192 Monte Carlo steps after discarding first 1,024 steps as
burn-in time.  By the jackknife analysis, the mean values and their
error bars (1$\sigma$) are finally evaluated as
\begin{align}
  \xi &= 1.040(7) \times 10^4 \\
  \Delta &= 7.99(5) \times 10^{-4}.
\end{align}
Since $\beta\Delta > 6$ and $L/\xi > 6$ are both satisfied, we can empirically regard the present
estimates as those in the thermodynamic and the
zero-temperature limits~\cite{TodoK2001,TodoMYT2001}.  The final estimate of $\Delta$ for $S=4$ is
plotted in Fig.~\ref{fig:gap} together with those for $S=1$, 2, 3
obtained by the previous quantum Monte Carlo
simulations~\cite{TodoK2001}.  It is clearly seen that $\Delta(S)$
decreases exponentially as $S$ increases.  The present result for the
gap can be compared with the lower and upper bounds for the gap
obtained by means of the numerical diagonalization~\cite{NakanoT2009}.
The present result is in between the lower and upper bounds, but it's
quite close to the upper bound.  On the other hand, the result for the
correlation length in the real-space direction is new to the best of
our knowledge.

It should be emphasized that the present results are obtained without
any extrapolation procedure; they are simply obtained by a single
Monte Carlo run on the largest system at the lowest temperature.
Thus, the highly parallelized continuous time loop algorithm is proved
to be a sensitive numerical tool that can distinguish a system with
extremely small excitation gap ($\sim 10^{-3}$) from a critical one,
and will certainly be necessary in investigating the properties of
novel quantum critical phenomena
in future theoretical studies in the statistical and condensed-matter
physics.

\section{Conclusion}
\label{sec:Conclusion}

In the present paper, we discuss the parallel quantum Monte Carlo
method that can simulate millions of spins at extremely low
temperatures ($\sim 10^{-6}$).  By using the nonlocal update scheme
based on the union-find graph algorithm, one can simulate such a huge
system without any convergence problem.  We show that the global graph
algorithm is performed very efficiently up to 24,576 nodes (196,608
cores) of the K computer by means of the OpenMP-MPI hybrid scheme
combined with several new techniques; asynchronous wait-free
union-find algorithm, butterfly-type global union-find algorithm,
majority-vote trick, process mapping optimization, etc.  In
conclusion, we have achieved $8 \,
(\text{threads}) \times 98.2\% \times 24,576 \, (\text{nodes}) \times
46.9\% \simeq 10^5$-fold speed-up by parallelization.  Together with
$\xi^2 \simeq 10^{8}$-fold acceleration of the
Monte Carlo dynamics by eliminating the critical slowing down by the
nonlocal cluster updates, we have virtually achieved $10^{13}$-fold
speed-up in total compared with the conventional local update quantum Monte
Carlo updates performed on a single core.
The present parallelization scheme can be extended to other models, such as the spin system with SU($N$) symmetry and the four-body interaction~\cite{HaradaSOMLWTK2013,SuzukiHMTK2015}, the transverse-field Ising model, etc.

\section*{Acknowledgment}

The authors would like to thank Tsuyoshi Okubo and Tatsuhiko Shirai for careful reading of the manuscript and comments.
The benchmark results presented in this paper have been obtained by
the K computer at the RIKEN Advanced Institute for
Computational Science.  The simulation program was developed based on
the ALPS library~\cite{ALPS2011,ALPSweb} and the ALPS/looper library~\cite{TodoK2001,LOOPERweb}.  This work was supported by
Grand Challenges in Next-Generation Integrated Nanoscience,
Next-Generation Supercomputer Project, the Strategic Programs for
Innovative Research (SPIRE), and a social and scientific priority issue (Creation of new functional devices and high-performance materials to support next-generation industries; CDMSI) to be tackled by using post-K computer from the MEXT, Japan.
S.T. acknowledges the support by KAKENHI (No.\,23540438, 26400384, 17K05564) from JSPS.

\bibliography{main}

\end{document}